\documentclass[aps,prl,twocolumn,superscriptaddress]{revtex4-1}

\usepackage{amsfonts}
\usepackage{amssymb}
\usepackage{amsmath}
\usepackage{upgreek}
\usepackage{color} 
\usepackage{lineno}
\newcommand{\tr}{\,\mathrm{tr}}
\newcommand{\myImag}{\mathrm{Im}}
\usepackage{braket}
\usepackage{mathtools}
\usepackage{ bbold }
\usepackage{graphicx}
\usepackage{dcolumn}
\usepackage{bm}

\usepackage[percent]{overpic}
\usepackage[export]{adjustbox}
\usepackage{stackengine}
\usepackage{appendix}

\newcommand\norm[1]{\left\lVert#1\right\rVert}

\usepackage{graphicx}
\usepackage{hyperref}
\hypersetup{colorlinks=true,
            citecolor=[rgb]{0,0,0.502},
            urlcolor=[rgb]{0,0,0.502},
            linkcolor=[rgb]{0,0,0.502}}
            
\newcommand{\beginsupplement} {
    \setcounter{table}{0}
    \renewcommand{\thetable}{S\arabic{table}}
    \setcounter{figure}{0}
    \renewcommand{\thefigure}{S\arabic{figure}}
    \setcounter{equation}{0}
    \renewcommand{\theequation}{S\arabic{equation}}
}

\newcommand*\patchAmsMathEnvironmentForLineno[1]{%
  \expandafter\let\csname old#1\expandafter\endcsname\csname #1\endcsname
  \expandafter\let\csname oldend#1\expandafter\endcsname\csname end#1\endcsname
  \renewenvironment{#1}%
     {\linenomath\csname old#1\endcsname}%
     {\csname oldend#1\endcsname\endlinenomath}}%
\newcommand*\patchBothAmsMathEnvironmentsForLineno[1]{%
  \patchAmsMathEnvironmentForLineno{#1}%
  \patchAmsMathEnvironmentForLineno{#1*}}%
\AtBeginDocument{%
\patchBothAmsMathEnvironmentsForLineno{equation}%
\patchBothAmsMathEnvironmentsForLineno{align}%
\patchBothAmsMathEnvironmentsForLineno{flalign}%
\patchBothAmsMathEnvironmentsForLineno{alignat}%
\patchBothAmsMathEnvironmentsForLineno{gather}%
\patchBothAmsMathEnvironmentsForLineno{multline}%
}

\newcommand{\mytitle}{Non-Gaussian correlations imprinted by local dephasing in fermionic wires}

\newcommand{\Mainz}{Institut f\"ur Physik, Johannes Gutenberg Universit\"at Mainz, D-55099 Mainz, Germany}
\newcommand{\Harvard}{Department of Physics, Harvard University, Cambridge, Massachusetts 02138, USA}
\newcommand{\UA}{Department of physics, Universiteit Antwerpen, B-2610 Antwerpen, Belgium}

\begin{document}

\title{\mytitle}

\author{Pavel~E.~Dolgirev}
\email[Correspondence to: ]{p\_dolgirev@g.harvard.edu}
\affiliation{\Harvard}
\author{Jamir~Marino}
\affiliation{\Harvard}
\affiliation{\Mainz}
\author{Dries~Sels}
\affiliation{\Harvard}
\affiliation{\UA}
\author{Eugene~Demler}
\affiliation{\Harvard}

\def\mean#1{\left< #1 \right>}

\date{\today}

\begin{abstract}
We study the behavior of an extended fermionic wire coupled to a local stochastic field. Since the quantum jump operator is Hermitian and quadratic in fermionic operators, it renders the model soluble, allowing investigation of the properties of the non-equilibrium steady-state and the role of dissipation-induced fluctuations. We derive a closed set of equations of motion solely for the two-point correlator; on the other hand, we find, surprisingly, that the many-body state exhibits non-Gaussian correlations. Density-density correlation function demonstrates a crossover from a regime of weak dissipation characterized by moderate heating and stimulated fluctuations to a quantum Zeno regime ruled by strong dissipation, which tames quantum fluctuations. Instances of soluble dissipative impurities represent an experimentally viable platform to understand the interplay between dissipation and Hamiltonian dynamics in many-body quantum systems.

\end{abstract}

\maketitle


The interplay of quantum many-body dynamics and decoherence is essential for understanding a broad range of physical phenomena: from suppression of weak localization of electrons due to coupling to phonons and electron-electron interactions~\cite{altshuler1998phase,gornyi2005interacting,basko2006metal,basko2008interplay}, to the realization of light-induced topological phases~\cite{mciver2012control,kitagawa2010topological,lindner2011floquet, oka2009photovoltaic,mciver2020light}, to the operation of quantum optical devices~\cite{walls2007quantum}, and to the implementation of quantum computers~\cite{preskill1998reliable} and simulators~\cite{daley2008quantum}. Competing effects of quantum entanglement and decoherence are also at the heart of questions of quantum control and quantum non-demolition measurements. One of the most surprising recent findings in this field is the effect of weak measurements on quantum fluctuations and on the decay rate of an excited state into a bosonic bath~\cite{caldeira1981influence,pichler2010nonequilibrium}. Depending on system parameters, this decay rate can be either inhibited or enhanced by weak measurements, with the two phenomena referred to as quantum Zeno and anti-Zeno effects~\cite{misra1977zeno,itano1990quantum,facchi2002quantum,kofman1996quantum, kofman2001zeno,kofman2000acceleration}, respectively.
\begin{figure}[b!]
\includegraphics[width=8.3cm]{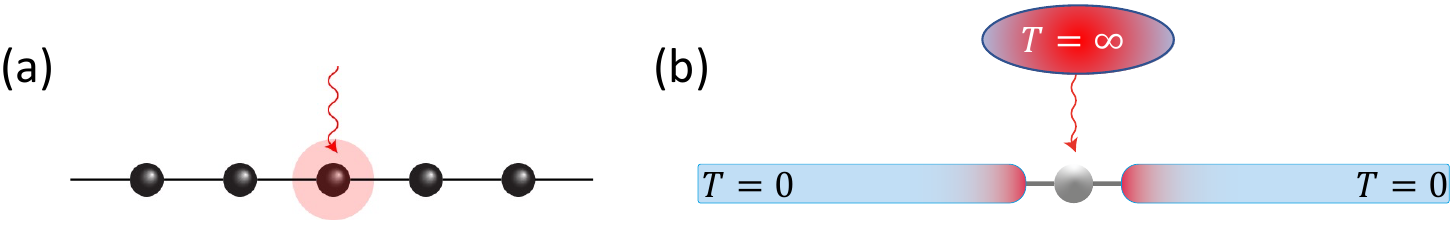}
 \caption{Sketch of the model. (a) A fermionic wire is stochastically driven by a local dephasing serving as continuous measurements (the strength $\gamma$ of dephasing is related to the measurement frequency~\cite{shchesnovich2010control}). (b) Alternatively, the model describes a quantum dot coupled to an infinite-temperature reservoir (embodying dephasing) and to zero-temperature leads with finite bandwidth ${\cal W}$ (the degrees of freedom of the wire). The Zeno physics manifests in the interplay between the bandwidth ${\cal W}$ and measurement frequency $\gamma$.}
\label{Setup}
\end{figure}
Several powerful techniques have been applied to the analysis of the interplay of quantum dynamics and decoherence in the many-body Zeno problem~\cite{elliott2015multipartite, mazzucchi2016quantum, cao2018entanglement, chan2019unitary, skinner2019measurement}, including Lindblad quantum master equations, memory kernel formalism, Keldysh diagrammatic techniques, and renormalization group approaches~\cite{vznidarivc2010exact,vzunkovivc2010explicit, dalla2012dynamics,prosen2014exact,Ott1, Ott2, Ott3, Ott4, krapivsky2019free, tonielli2019orthogonality, wasak2019quantum, corman2019quantized, PhysRevLett.116.235302, berdanier2019universal, wolff2019evolution, froml2019ultracold,froml2019fluctuation, goold}. Examples of quantum many-body systems with decoherence that allow exact theoretical solutions and can be realized experimentally are particularly valuable, since they enable a non-perturbative analysis of competing effects of dissipation. So far, such systems -- including Bethe ansatz solution~\cite{ziolkowska2019yang} of noisy tight-binding fermions~\cite{PhysRevLett.117.137202}, boundary driven quantum spin chains~\cite{prosen2015matrix,buca}, and non-hermitian Richardson-Gaudin magnets~\cite{PhysRevLett.120.090401}  -- have been few and far between. On the experimental side, modern solid-state and cold-atom platforms---including atomic BEC with local losses, disordered trapped ion strings with dissipation facilitated transport properties, and the realization of dissipative scanning-gate microscopes with $^6$Li atoms---already enable investigation of effects of dissipation both in the form of losses and dephasing~\cite{syassen2008strong,Ott3,Ott4,PhysRevLett.115.083601,corman2019quantized,lebrat2019quantized,PhysRevLett.122.050501}.

In this work, we present a new example of such a system, in which a local stochastic field couples to the electron density on a single site in a one-dimensional fermionic chain (see Fig.~\ref{Setup}). This system has recently been realized experimentally~\cite{corman2019quantized}. Here we provide a theoretical analysis of this model and make predictions which can be tested with currently available experimental platforms.

Before entering  the technical details of our work, we provide an overview of the key results. When a stochastic field couples locally to the electron density, it introduces two opposing effects on fluctuations in the difference of the number of particles between the left and right parts of the fermionic chain. The stochastic field provides local heating and thus enhances fluctuations; on the other hand, it performs `weak measurements' of the electron number, which hinders particle propagation across the site with decoherence, suppressing relative number fluctuations. We find that the competition between these two effects leads to the existence of two distinct regimes of dynamics: for weak dissipation, number fluctuations become enhanced with increasing decoherence; in contrast, for strong dissipation, fluctuations become suppressed. The two regimes are separated by a sharp crossover displayed in Fig.~\ref{SSE}b,c. A special feature of the local dephasing problem from the mathematical viewpoint is that the BBGKY hierarchy is closed, meaning that the equation of motion for a given $n$-point correlation function can be expressed through correlators whose order is $n$ or less -- for instance, Eq.~\eqref{eqn::main} represents the evolution of the two-body correlation function. Such a situation typically occurs for Gaussian systems~\cite{shi2018variational}, where high-order correlators factorize in terms of the two-body correlators. However, one does not expect the state to be Gaussian. Indeed, for a given realization of the fluctuating field, the state is Gaussian since the system evolves under a non-interacting Hamiltonian with time-dependent stochastic potential, while after averaging over different realizations of the noise, the state becomes non-Gaussian~\cite{cucchietti2005decoherence}. Figure~\ref{SSE}b further supports such conclusion. This remarkable circumstance that the BBGKY hierarchy is closed, despite the emergence of non-Gaussian correlations, allows us to investigate the effects of dissipation non-perturbatively.

\begin{figure}[t!]
\includegraphics[width=8.5cm]{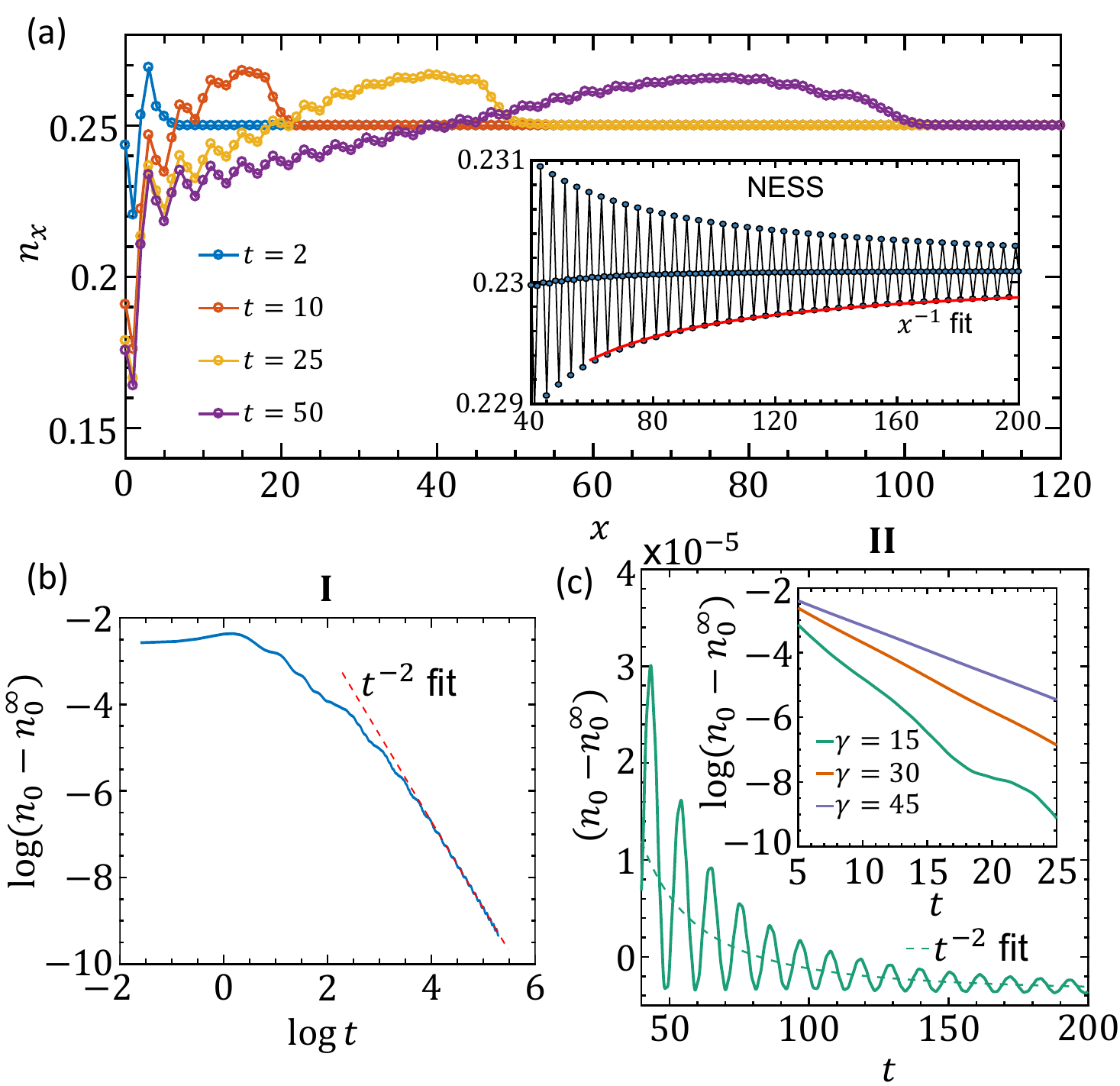}
 \caption{ Numerical simulation of Eq.~\eqref{eqn::main}. (a) Density profile $n(x,t)\equiv\Gamma_{xx}(t)$ at different times as a function of $x$. Note a ballistic density front emitted due to the dissipative impurity located at the origin. Here we fix $\gamma = J$. The inset shows the analytical prediction for the algebraic decay of spatial correlations when the system reaches the steady-state. (b) and (c): Fermionic density at the origin relaxes as $t^{-2}$ towards the NESS. In the regime of strong dissipation (II), dynamics is exponentially damped at short times (inset of panel (c)), and has a superimposed oscillatory behaviour at late times. } 
\label{Lindblad}
\end{figure}

\begin{figure}[t!]
\includegraphics[width=8.5cm]{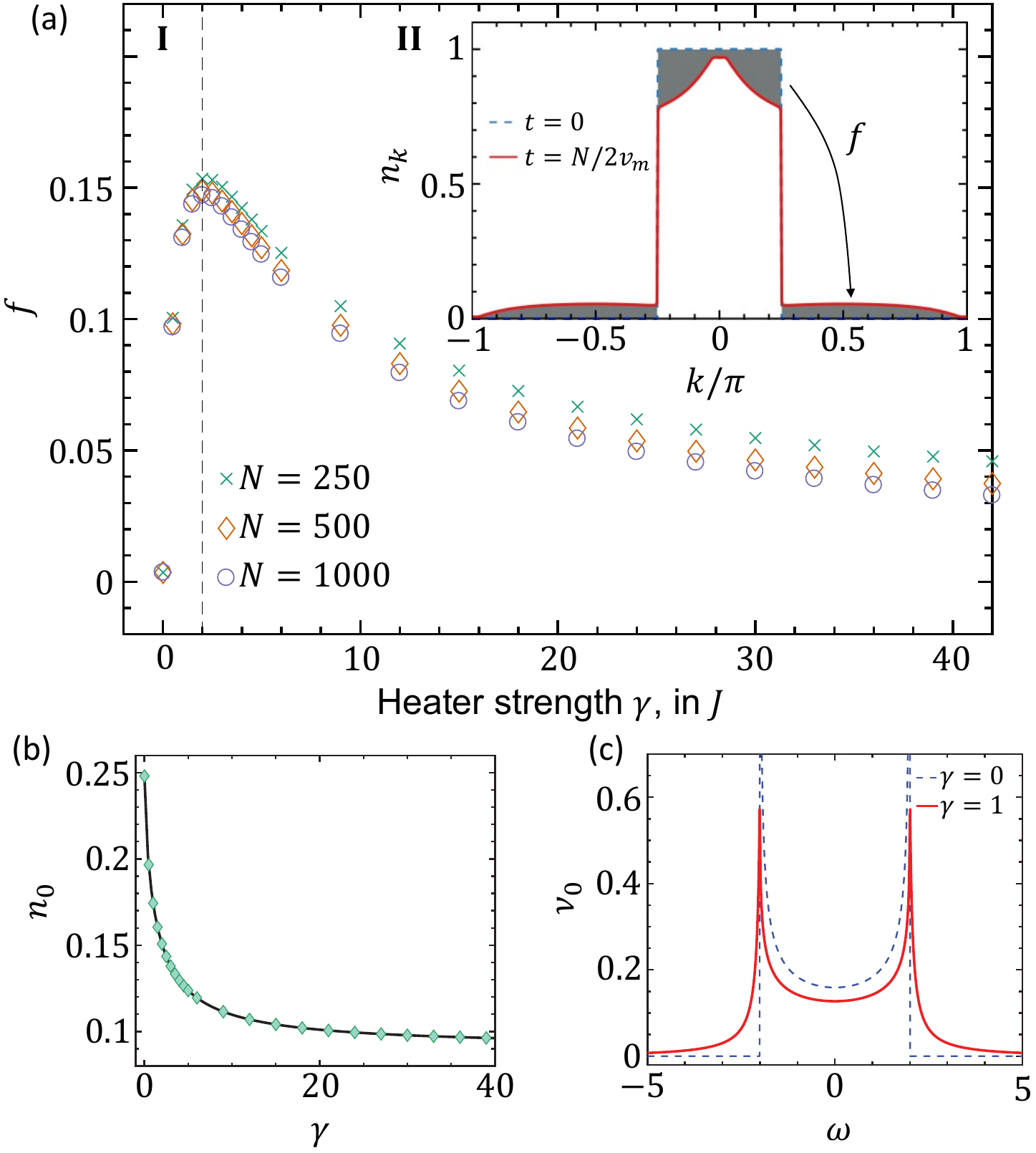}
 \caption{(a) Fraction of fermions left inside the Fermi pocket, $f$, evaluated at longest simulation times, as a function of the dissipation strength $\gamma$, demonstrating the Zeno crossover. Inset: momentum distribution of fermions $n_k \equiv \Gamma_{kk}$ at $t = 0$ and at $t = N/2v_m$, showing that the heater results in the redistribution of fermions towards high-momenta states; note that the Fermi edge (we fix $k_F = \pi/4$) is preserved. (b) A finite density of fermions at $x=0$ remains in the NESS even at strong dissipative rates; the numerical solution (diamonds) is on top of the analytical prediction (solid line). (c) Local DOS at site $x=0$ in the NESS as a function of frequency. }
\label{NESS}
\end{figure}

Throughout the paper, we develop several complementary approaches for solving the model in Fig.~\ref{Setup}. Let us start with the formalism of the quantum master equation (QME). The fermionic wire is described with the following Hamiltonian:
\begin{align}
    \hat{H}_0 = \sum_k \xi_k {\hat c}^\dagger_{k}{\hat c}_{k},\, \xi_k=-2J\cos k - \mu, \label{eqn::Hamiltonian}
\end{align}
The operator $c_k^{(\dag)}$ annihilates (creates) a fermion with momentum $k$ (the distance between neighboring sites is set to unity). $\mu$ is chemical potential. $J$ is the hopping between nearby sites and it sets the unit of energy. The Planck constant $\hbar = 1$ throughout the paper. The density matrix evolves according to (QME)
\begin{equation}
    \frac{d \hat{\rho}}{dt} = -i [\hat{H}_0,\hat{\rho}] + \gamma (\hat{L}\hat{\rho}\hat{L}^\dagger - \frac{1}{2}\{\hat{L}^\dagger\hat{L},\hat{\rho}  \}), \label{eqn::rho_QME}
\end{equation}
where $\hat{L} = \hat{n}_{0} = \frac{1}{N}\sum_{p,q}{\hat c}^\dagger_{p}{\hat c}_{q}$ is the quantum jump operator representing the local dephasing. It is worth emphasizing that, although the Hamiltonian~\eqref{eqn::Hamiltonian} is non-interacting, the complexity of the underlying evolution is due to local quartic `dissipative interactions' arising from the second term in Eq.~\eqref{eqn::rho_QME}. Significant simplification occurs by noting that $\hat{L}^\dagger = \hat{L}$ and the dynamics of any observable $\hat{\cal O}$ can be written as: $$\frac{d}{dt} {\cal O} = \frac{d}{dt}\tr {\hat{\cal O} \hat{\rho}} = i\mean{[\hat{H}_0,\hat{\cal O}]} + \frac{\gamma}{2}\mean{[\hat{L},[\hat{\cal O},\hat{L}]]}.$$ As long as the Hamiltonian and the quantum jump operator are both quadratic in the fermionic operators, it follows that the evolution of any $n$-point correlation function $\Gamma^{k_1,k_2,\dots,k_n}_{k'_1,k'_2,\dots, k'_n}(t) \equiv \langle \hat{c}^\dagger_{k_1}\hat{c}_{k'_1} \dots \hat{c}^\dagger_{k_n}\hat{c}_{k'_n} \rangle$ can be expressed through operators whose order is $n$ or less; in other words, we can write down a closed system of equations of motion for any given order of interest~\cite{haken1973exactly}. In particular, for $\Gamma_{k k'}(t)\equiv \langle\hat{c}^\dagger_{k}\hat{c}_{k'}\rangle$ we get
\begin{eqnarray}
\frac{d}{dt}\Gamma_{k k'}(t) = i(\xi_k - \xi_{k'})\Gamma_{k k'} + \frac{\gamma}{N^2}\sum_{p,q} \Gamma_{p,q} \notag{}\\
-\frac{\gamma}{2N} \sum_{q}(\Gamma_{k,q}+\Gamma_{q,k'}). \label{eqn::main}
\end{eqnarray}
Note that the total number of fermions $N_{\rm tot} = \sum_k \Gamma_{kk}$ is conserved, $\dot{N}_{\rm tot} = 0$. Although we are able to write down a closed equation of motion for the two-body correlation function, we find that the many-body density matrix is genuinely non-Gaussian: even if the initial state, such as a filled Fermi sea, is Gaussian, the local dephasing will imprint non-Gaussian correlations. A possible way to see this is to investigate higher-order correlation functions. For example, in Ref.~\cite{SM} we demonstrate, by explicitly deriving the corresponding equation of motion, that the four-point correlator cannot be factorized (using Wick's theorem) in terms of the two-body correlation functions. The mentioned equation turns out to be a challenge for numerical treatments of extended systems, motivating the development of an alternative approach for investigating higher-order correlators below.

We turn to explore Eq.~\eqref{eqn::main} both numerically and analytically. In Fig.~\ref{Lindblad}, we plot the evolution of the fermionic density profile $n_x(t)$: the heater locally perturbs the system, resulting in the emission of a ballistic density front (its velocity equals to the maximum group velocity $v_m = 2J$); at the same time, in the vicinity of the dissipative impurity, a non-equilibrium steady-state (NESS) forms~\cite{NESS}. 
To investigate NESS properties numerically, we let the system evolve up to times $\sim N/2v_m$. In momentum space, since the total energy is not conserved, we find that the initial Fermi sea becomes redistributed towards large-momenta states, as shown in the inset of Fig.~\ref{NESS}a. Interestingly, the system preserves the Fermi edge, and its distribution function becomes non-thermal. We characterize the cumulative effect of the heater by the fraction $f$ of fermions removed from the Fermi pocket. Figure~\ref{NESS}a shows that $f$, evaluated at the longest simulation time $t = N/2v_m$,  exhibits a crossover as a function of $\gamma$, switching from the anti-Zeno regime at week dissipation to the Zeno regime at strong dissipation.

We now discuss the approach towards the NESS. As shown in Fig.~\ref{Lindblad}b,c, our numerical analysis indicates that in both regimes of weak (I) and strong (II) dissipation, the system exhibits $t^{-2}$ relaxation (on top of oscillatory behavior) -- see~\cite{SM} for further details. Qualitative differences are present in the intermediate-time dynamics, as one can inspect from Fig.~\ref{Lindblad}b,c: for stronger dissipation, the density at the origin demonstrates slower evolution. Indeed, strong noise drives out of resonance hopping processes involving the dissipative site, implying a `trapping' of particles jumping to the origin and a suppression of transport across this site.

Although the dynamics encoded in Eq.~\eqref{eqn::main} can be efficiently  simulated numerically, they represent a challenge for analytical solutions. Remarkably, diagrammatic field theory provides an alternative derivation of Eq.~\eqref{eqn::main}, and allows to extract analytically the NESS properties.
We start by noticing that the Linbladian evolution in Eq.~\eqref{eqn::rho_QME} is equivalent to the stochastic Schrodinger equation (SSE) with Hamiltonian:
\begin{equation}
    \hat{H}_{\xi}(t) = \hat{H}_0 + \hat{V}(t), \hat{V}(t) = \xi(t) \hat{n}_{0}, \label{eqn::SSE}
\end{equation}
where $\xi(t)$ is a white noise with $\mean{\xi(t_1)\xi(t_2)}_{\xi} = \gamma\delta(t_1-t_2)$. To compute the dynamics of any observable $\hat{\cal O}$, one needs to perform averaging over the noise:
 $$\displaystyle {\cal O}(t) = \mean{{\cal O}_{\xi}(t) }_{\xi}, {\cal O}_{\xi} = \tr \left(\hat{{\cal O}} \hat{\rho}_{\xi} \right),$$ 
where $\hat{\rho}_{\xi}$ is the density matrix for a given noise realization $\xi(t)$. The conservation of the total number of fermions follows from $[\hat{N}_{\rm tot},\hat{H}_{\xi}(t)] = 0$. The initial density matrix at $t = 0$ is chosen to be a filled Fermi sea.

\begin{figure}[t!]
\includegraphics[width=8.3cm]{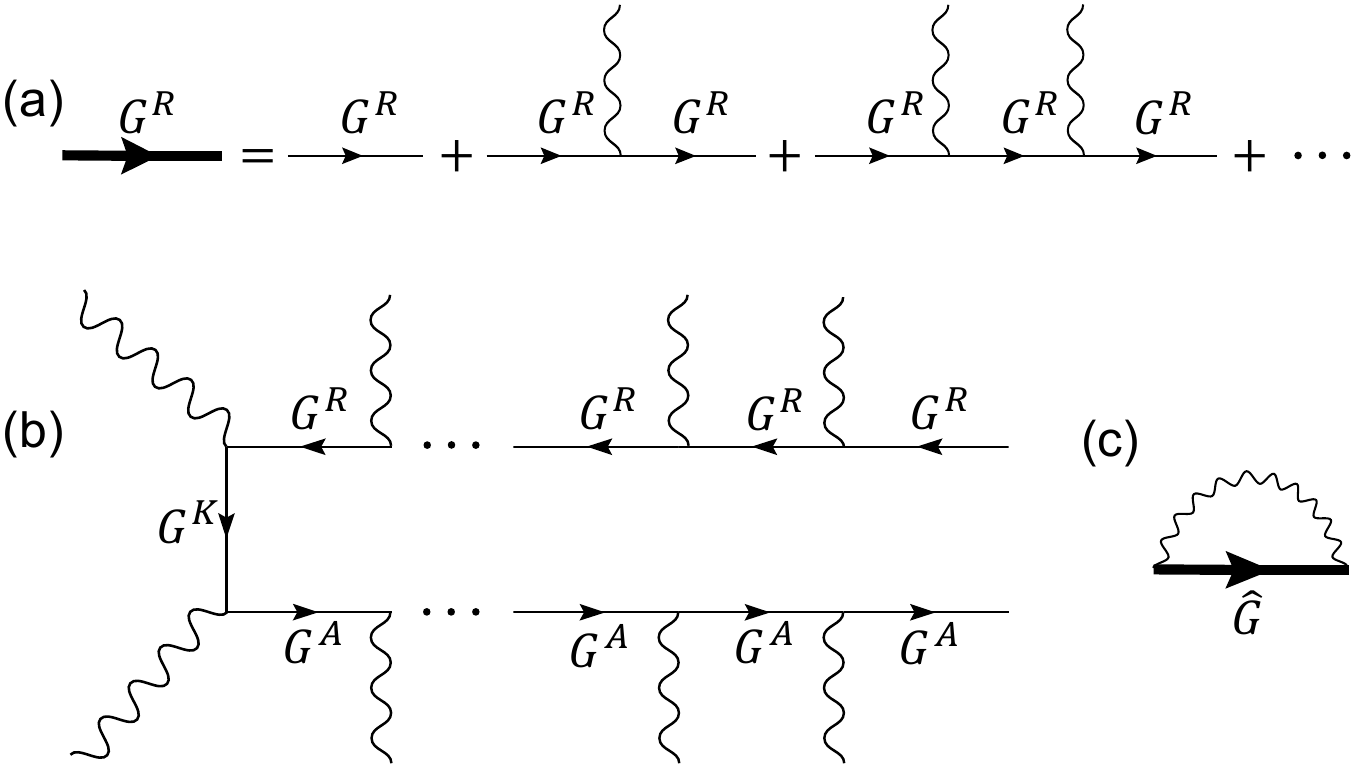}
 \caption{(a) Schematics of the Dyson series for the retarded Green's function, cf. Eq.~\eqref{GR}; and (b) for the Keldysh Green's function, cf. Eq.~\eqref{GK}. (c) The  self-energy obatined by averaging over the noise the Dyson equations~\eqref{GR} and~\eqref{GK}.}
\label{Diag}
\end{figure}

We now develop a non-equilibrium diagrammatic technique inspired by the treatment of disordered fermionic systems~\cite{kamenev2011field}. For the retarded Green's function, defined as $G^R_{tt'}(k,k') \equiv -i \theta(t-t')\mean{  \left\{ \hat{c}_k(t), \hat{c}^{\dagger}_{k'}(t') \right\} }$, the  Dyson series~\cite{kamenev2011field} reads (see Fig.~\ref{Diag}a):
\begin{align}\label{GR}
    \hat{G}^R= \sum_{m,n=0}^{\infty}  (\hat{G}_0^R\circ \hat{V})^m \circ \hat{G}^R_0 \circ(\hat{V}\circ \hat{G}_0^R)^n.
\end{align}
Here $\hat{G}_0^R$ is the unperturbed retarded Green's function. Because the underlying problem is far from equilibrium, we will also need to compute the Keldysh Green's function, $G^K_{tt'}(k,k') \equiv -i \mean{ \left[ \hat{c}_k(t), \hat{c}^{\dagger}_{k'}(t') \right] }$:
\begin{align}\label{GK}
    \hat{G}^K = \sum_{m,n=0}^{\infty}  (\hat{G}_0^R\circ \hat{V})^m \circ \hat{G}^K_0 \circ(\hat{V}\circ \hat{G}_0^A)^n.
\end{align}
An element of this series is schematically depicted in Fig.~\ref{Diag}b. Because the noise is local in space and time, averaging results in the self-energy known as a self-consistent Born approximation (SCBA) -- shown in Fig.~\ref{Diag}c, which in our case holds exactly. For further technical details we refer to Ref.~\cite{SM}, where, in particular, we show that the equation for the equal-time Keldysh Green's function reduces to Eq.~\eqref{eqn::main}.

 \begin{figure}[t!]
\includegraphics[width=8.cm]{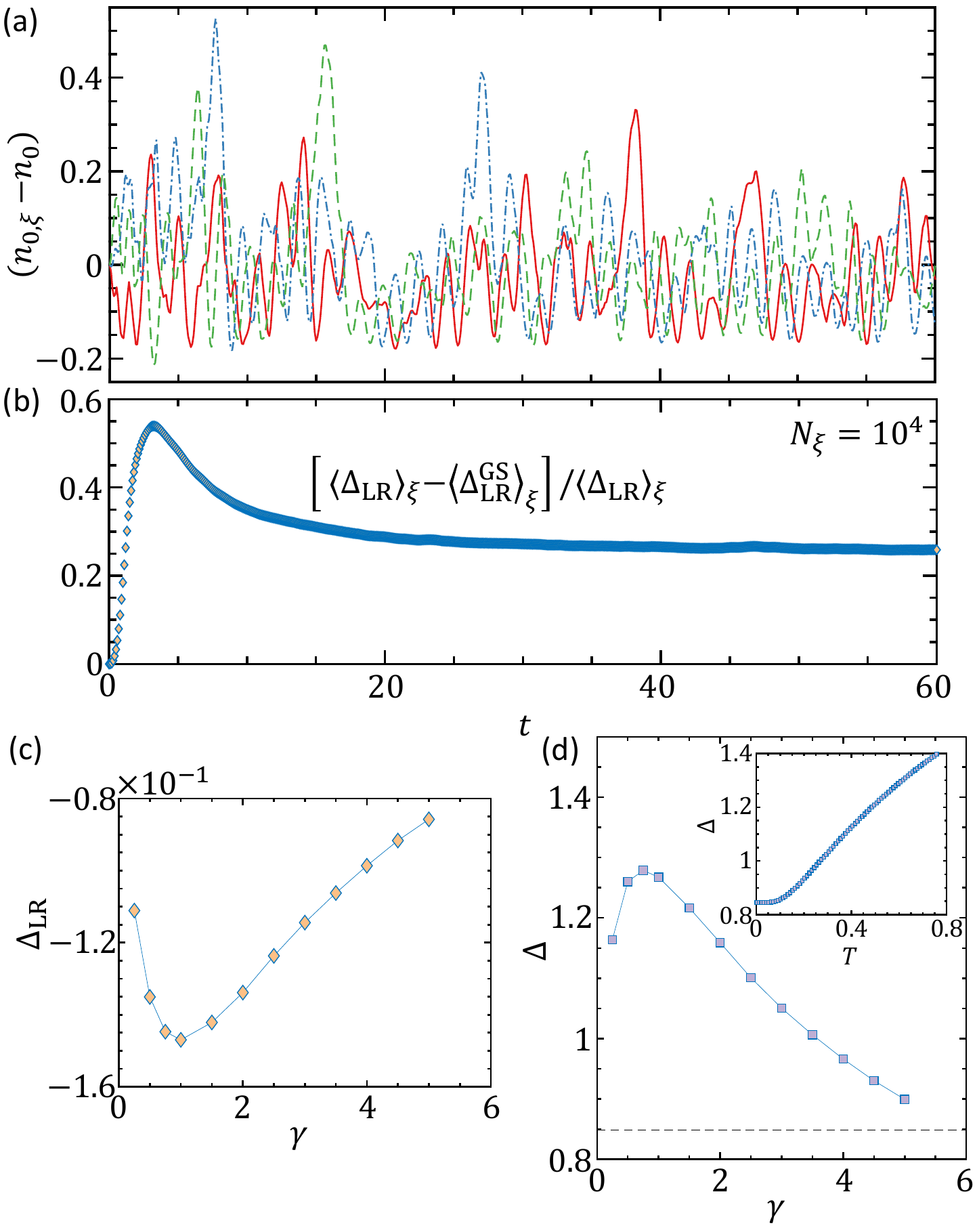}
 \caption{Numerical simulations of the SSE. (a) Evolution of the density at the origin $n_{0,\xi}(t)$ for three different noise realizations ($N = 200$, $\gamma = J$, and $k_F = \pi/4$). We find that the typical instance of dynamics is far from the average, $n_{0}(t)$. (b)  Deviation of the noise averaged correlator $\langle\Delta^\xi_{\rm LR}\rangle_\xi$, cf. Eq.~\eqref{eqn:Delta}, from its  expectation value assuming a Gaussian state, demonstrating the development of non-Gaussian corrections. (c) and (d): Two fluctuation correlation functions, $\Delta_{\rm LR}$ and $\Delta$, exhibit the Zeno crossover in the NESS. Standard deviation in both observables is less than the `symbol' size. Inset of (c): the  dependence of $\Delta$ on temperature is monotonous in equilibrium, further indicating that the heating effect is suppressed at strong dissipation.} 
\label{SSE}
\end{figure}

From~\eqref{GR}, we compute the NESS retarded Green's function in the frequency domain [Eq.~(S12) in Ref.~\cite{SM}], allowing to extract, for example, the local density of states (DOS) at the impurity site $\nu_0(\omega) \equiv -\frac{1}{\pi}\myImag \, G^R_\omega(0,0)$ -- see Fig.~\ref{NESS}c. For $\gamma\neq 0$ it develops tails at high frequencies; the presence of low-energy modes confirms the aforementioned power-law dynamics in the nearby of the NESS, see Fig.~\ref{Lindblad}b,c. This structure of the DOS can be directly measured in state-of-art solid-state experiments, which can access the dynamics of non-equilibrium quantum impurities~\cite{latta2011quantum}. 
Similarly, from Eq.~\eqref{GK}, we derive an expression for the NESS Keldysh Green's function [Eqs.~(S15) and~(S17) in Ref.~\cite{SM}], from which we compute, for instance, the spatial profile of the fermionic density. Figure~\ref{NESS}b shows that the analytical expression for the density at the origin, $n_0^\infty$, is in a remarkable agreement with the result numerically computed from~\eqref{eqn::main}. For $k_F = \pi/4$ ($k_F = 3\pi/4$), $n_0^\infty$ is a monotonically decreasing (increasing) function of $\gamma$, and it remains finite even for very strong dissipation. It is compelling that different single-body observables, such as $n_0^\infty$ and $f$, exhibit qualitatively distinct behavior -- see Fig.~\ref{NESS}a,b. Although dissipation occurs locally in space, we find that in the NESS, the fermionic density profile demonstrates long-range behavior following a~$x^{-1}$ fit with superimposed Friedel's oscillations -- see inset of Fig.~\ref{Lindblad}a. These two features are related to the fact that the system preserves the Fermi-edge singularity, shown in the inset of Fig.~\ref{NESS}a. This intrusive effect reminds the situation of a static impurity~\cite{nozieres1969singularities,mahan2013many}, and might be of relevance for experimental manipulations of quantum many-body systems subject to local dissipation.

We now turn to numerical simulation of the SSE which offers a complementary physical viewpoint. An infinitesimal time step is performed via `Trottorizing' the evolution operator: $\hat{U}^\xi_{t+\delta t, t} \approx {\rm e}^{-\frac{i}{2} \delta t \hat{H}_0} {\rm e}^{- i \delta W \hat{n}_0 } {\rm e}^{-\frac{i}{2} \delta t \hat{H}_0}$ with a time step $\delta t \ll \min\{\gamma^{-1},J^{-1}\}$. By $\delta W$ we denote a Wiener process~\cite{gardiner2004quantum} with $\mean{\delta W\delta W} = \gamma \delta t$.~Figure~\ref{SSE}a shows that the typical evolution of the density at the origin exhibits pronounced fluctuations and is far from the Linbladian result obtained from Eq.~\eqref{eqn::main}. Only after averaging over many ($N_\xi \sim 10^4$ for $\gamma = J$) noise realizations the two approaches start to match~\cite{SM}. This physical picture suggests that these pronounced fluctuations dominate the aforementioned algebraic behavior and non-equilibrium crossover.

Simulations of the SSE allow direct investigation of correlation functions associated with density fluctuations -- a formidable task for both the QME and diagrammatic methods. Specifically, we study two correlators:
\begin{align}
    \Delta^\xi_{\rm LR}\equiv \langle\hat{N}_L\hat{N}_R\rangle - \langle\hat{N}_L\rangle\langle\hat{N}_R\rangle,\, \Delta^\xi \equiv \langle(\hat{N}_L-\hat{N}_R)^2\rangle. \label{eqn:Delta}
\end{align}
$\hat{N}_R \equiv \sum_{i = 1}^{l}\hat{n}_{i}$ is the total density of fermions on $l$ sites on the  right of the impurity (we fix $l=5$). Analogously, we define the total density of fermions on the left, $\hat{N}_L$. Figure~\ref{SSE}b shows the difference between the dynamics of the noise-averaged correlator $\langle \Delta_{\rm LR}^\xi\rangle_\xi$ and  the evolution of the same quantity  assuming the system in a  Gaussian state. We find that non-Gaussian correlations, imprinted by the local dephasing channel, are strongest at intermediate times, when the system is already far from the filled Fermi sea, but didn't reach yet the NESS, which turns out also to be non-Gaussian (a similar conclusion for the problem of global dephasing is discussed in Ref.~\cite{vznidarivc2011solvable}). The non-Gaussian correlations are also more pronounced near the site with decoherence, as we show in Ref.~\cite{SM}. Deviations from Gaussianity are at reach in state-of-art cold-atoms experiments, as recently demonstrated in the measurement of higher-point correlation functions of phase profiles in a pair of tunnel-coupled one-dimensional atomic superfluids~\cite{schweigler2017experimental}.

We now focus on  NESS properties. At equilibrium, these correlators depend monotonically on temperature -- see inset of Fig.~\ref{SSE}d. After a relatively short time ($t_0 \approx 30J^{-1}$ for $\gamma = J$ -- see Fig.~\ref{SSE}b), both observables $\Delta^\xi_{\rm LR}$ and $\Delta^\xi$ demonstrate saturation, indicating proximity to the NESS. This fact suggests that by averaging over both time and noise, $
    \Delta_{(\rm LR)} \equiv \displaystyle\frac{1}{T}\int\limits_{t_0}^{t_0 + T} dt \langle\Delta^\xi_{(\rm LR)}\rangle_\xi,$
one probes the NESS correlations. Here we fix $N_\xi = 10^3$ and $T = 60 J^{-1}$. Figure~\ref{SSE}c,d shows that in the NESS such correlations exhibit a crossover around $\gamma = J$: For small $\gamma \lesssim J$, the equilibrium cartoon in the inset of Fig.~\ref{SSE}d suggests that the temperature of the system increases with the heater strength, in contrast to the case of strong $\gamma \gtrsim J$, where it starts to decrease. This inability to heat up the system for strong dissipation is the essence of the quantum Zeno effect. Our procedure also suggests the feasibility of experimental verification of this result (note that it is already accessible with cold-atom platforms~\cite{Ott4} to probe correlations after $\sim 20 J^{-1}$).

The Lindblad QME illustrates that the BBGKY hierarchy  is closed, which mirrors  at a diagrammatic level in the  exactness of the SCBA. This circumstance allows to extract analytically the NESS properties  and, in particular, to demonstrate the onset of algebraic spatio-temporal correlations. On the other hand, the simulations of the SSE enable to investigate the effects of fluctuations non-perturbatively.  We have found that these fluctuations exhibit non-Gaussian correlations and behave non-monotonically  as a function of the dissipation strength (this manifests in the Zeno crossover discussed thorough the paper). It would be interesting to extend this program to interacting systems both of fermionic or bosonic nature, in view of applications to  solid-state and cold-atoms experiments. As an example, the question of fluctuations statistics in the setup of coupled Josephson-junction arrays with local dissipation~\cite{PhysRevLett.116.235302} is particularly promising.

 \begin{acknowledgments}
We thank J. Schmiedmayer, P.L. Krapivsky, T. Esslinger, I. Kukuljan, M.H. Michael, G. Zar{\'a}nd, D. Abanin, and R. Schilling for fruitful discussions. P.E.D. and E.D. are supported by the Harvard-MIT Center of Ultracold Atoms, AFOSR-MURI Photonic Quantum Matter (award FA95501610323), and DARPA DRINQS program (award D18AC00014). J.M. was supported by the European Union's Horizon 2020 research and innovation programme under the Marie Sklodowska-Curie grant agreement No 745608 (QUAKE4PRELIMAT). D.S. acknowledges support from the FWO as post-doctoral fellow of the Research Foundation  Flanders.
\end{acknowledgments}

\bibliography{impurity_lib}

\begin{thebibliography}{66}
\bibitem [{NES()}]{NESS}%
  \BibitemOpen
  \href@noop {} {}\bibinfo {note} {We assume that the system is infinite; hence
  energy produced locally by the dephasing channel is carried away to spatial
  infinity, without revivals. The steady-state is defined only with respect to
  the nearby of the dissipative site, while far away from it the `density
  front' keeps spreading along the system.}\BibitemShut {Stop}%
\end{thebibliography}

\clearpage
\newpage

\beginsupplement

\section{Non-equilibrium diagrammatic approach: equations of motion}

\subsection{Retarded Green's function}

\begin{figure}[b!]
\centering
\includegraphics[width=1\linewidth]{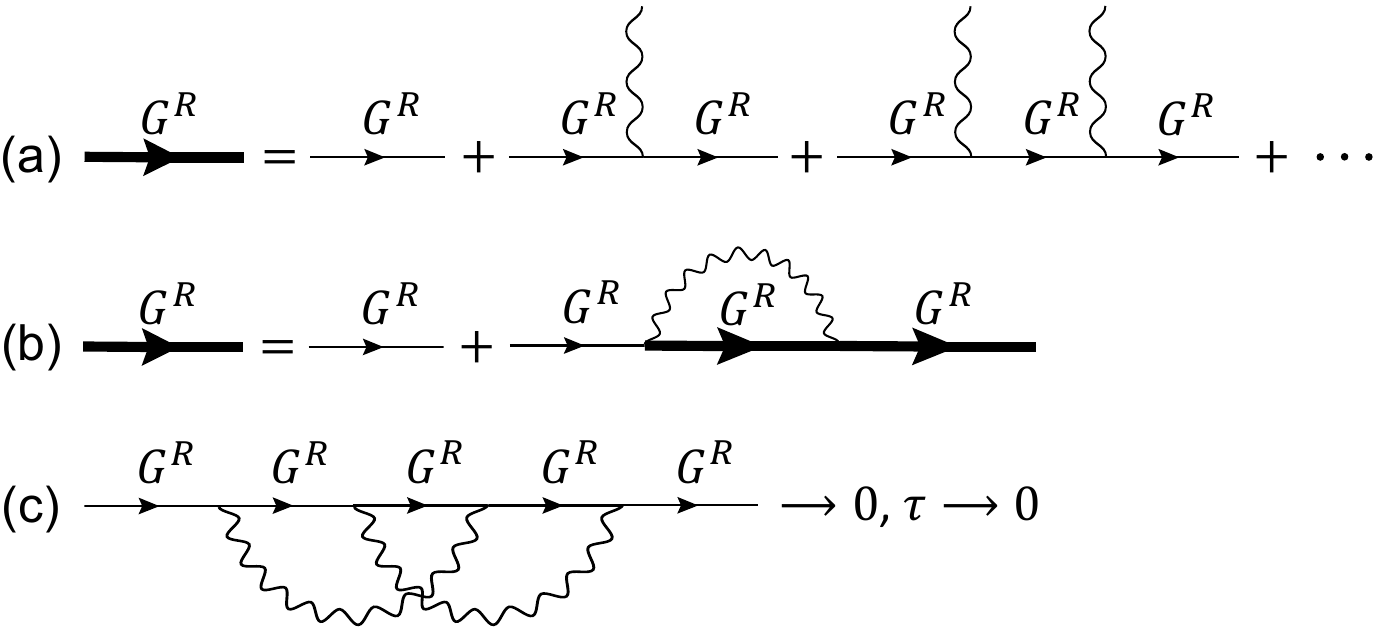} 
\caption{(a) The Dyson series for the retarded Green's function, cf. Eq.~\eqref{eqn:G_R}. (b) Result of averaging over the noise in (a) gives the SCBA. (c) A diagram with a crossing is proportional to $\tau$ and, thus, vanishes ($\tau\to0$).}
\label{fig::Retarded}
\end{figure}

The retarded Green's function is defined as 
\begin{align}
    G^R_{tt'}(k,k') \equiv -i \theta(t-t')\mean{ \left\{ \hat{c}_k(t), \hat{c}^{\dagger}_{k'}(t') \right\} }.\label{eqn::def_GR}
\end{align}
Expanding Eq.~\eqref{eqn::def_GR} perturbatively in powers of $\hat{V}$, we obtain the following Dyson series (cf. Fig.~\ref{fig::Retarded})
\begin{align}
    \hat{G}^R = \sum_{m,n=0}^{\infty}  (\hat{G}_0^R\circ \hat{V})^m \circ \hat{G}^R_0 \circ(\hat{V}\circ \hat{G}_0^R)^n, \label{eqn:G_R}
\end{align}
where $\hat{G}_0^R$ is the Green's function of the unperturbed Hamiltonian $\hat{H}_0$. Upon averaging Eq.~\eqref{eqn:G_R} over the noise, one typically introduces a  non-crossing approximation (the self-consistent Born approximation [SCBA], see Ref.~[\onlinecite{altland2010condensed}]), which is schematically illustrated in Fig~\ref{fig::Retarded}b. In our case, the SCBA holds exactly. We first explicitly evaluate the resulting self-energy

\begin{equation*}
    \Sigma^R_{tt'}(k,k') = \gamma  \delta(t-t') \frac{1}{N^2}\sum_{p,q} G^R_{t,t}(p,q) \times \theta(t-t_0),
\end{equation*}
where $t_0$ is the time at which  dissipation has been switched on. One immediately recognizes that the equal-time retarded Green's function is ill-defined because of the Heaviside $\theta$-function in Eq.~\eqref{eqn::def_GR}: $G^R_{tt}(k,k') = -i \theta(0) \delta_{kk'}$. This problem originates from the choice of the $\delta$-correlated noise in Eq.~(5). We consider instead a non-Markovian bath with $\mean{\xi_t\xi_{t'}} = \gamma f_{\tau}(t-t')$, where $f_{\tau}$ is a bell-shaped symmetric smooth function approximating the $\delta$-function: $\lim_{\tau\rightarrow0} f_{\tau}(t-t')=\delta(t-t')$ and $\int f_{\tau }(t) dt = 1$. For this choice, the SCBA gives the leading order in $\tau$ contribution
\begin{align}
    \Sigma^R_{tt'}(k,k') \approx \gamma & \delta(t-t')\theta(t-t_0)\notag{} \\
    & \times\lim_{\tau\rightarrow 0} \frac{1}{N^2}\sum_{p,q}\int d t''  G^R_{t,t''}(p,q) f_{\tau}(t - t'') .\notag{}
\end{align}
Note that $f_\tau (t-t'')$ restricts $t''$ to be in the $\tau$-vicinity of $t$, and the retarded Green's function further limits $t'' \leq t$, i.e. only a ``half'' of $f(t-t'')$ contributes to $\Sigma^R$. Then, the limit $\tau\to 0$ gives
\begin{equation}
    \Sigma^R_{tt'}(k,k') =    \frac{-i\gamma}{2N} \delta(t-t') \times \theta(t-t_0).\label{eqn::Sigma_R_SCBA}
\end{equation}
From this expression, we conclude that the proper choice for defining the $\theta$-function corresponds to $\theta(0) = \frac{1}{2}$.

Diagrams with crossings vanish, see Fig.~\ref{fig::Retarded}c for an instance. A diagram of this class contains a product $G^R_{t_1t_2} G^R_{t_2t_1}$, which may be non-zero only for $t_1=t_2$, which is a manifold of zero measure. Alternatively, one can observe that this diagram vanishes in the limit $\tau\to 0$.

\subsection{Keldysh Green's function}
\begin{figure}[hbt!]
\centering
\includegraphics[width=0.7\linewidth]{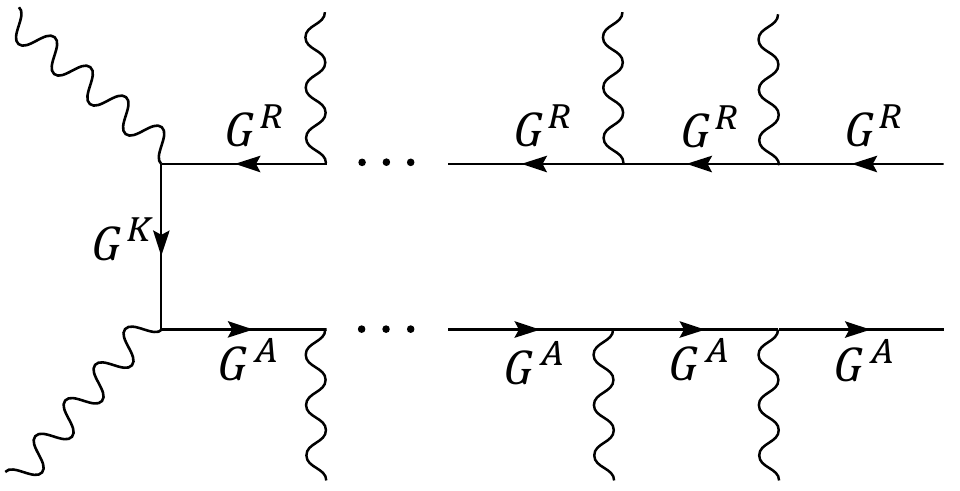} 
\caption{Structure of diagrams entering the Dyson series for the Keldysh Green's function, cf Eq.~\eqref{eqn:G_K}.}
\label{fig::Keldysh}
\end{figure}

The Dyson series for the Keldysh component reads (cf.  Fig.~\ref{fig::Keldysh})
\begin{align}
    \hat{G}^K = & \sum_{m,n=0}^{\infty}  (\hat{G}_0^R\circ \hat{V})^m \circ \hat{G}^K_0 \circ(\hat{V}\circ \hat{G}_0^A)^n,  \label{eqn:G_K}
\end{align}
where $\hat{G}^K_0$ is the Keldysh Green's function of the unperturbed Hamiltonian $\hat{H}_0$. Using the result of the previous subsection, the series~\eqref{eqn:G_K} can be rewritten as
\begin{align*}
    \hat{G}^K = & \, \hat{G}^K_0 + \hat{G}^R\circ \hat{V} \circ \hat{G}^K_0 \circ\hat{V}\circ \hat{G}^A  \notag{} \\
    &\quad + \hat{G}^R\circ \hat{V} \circ \hat{G}^K_0 + \hat{G}^K_0 \circ\hat{V}\circ \hat{G}^A.
\end{align*}
From this latter expression we deduce that the Keldysh component of the resulting self-energy is given by
\begin{equation}
    \Sigma^{K}_{t_1t_2}(k,k') = \frac{\gamma}{N^2} \delta(t_1-t_2) \theta(t_1-t_0) \sum_{p q} G^K_{t_1t_1}(p,q). \label{eqn::Sigma_K_SCBA}
\end{equation}

\subsection{Equations of motion}
From the Dyson equations: $(\hat{G}^{-1}_0 - \hat{\Sigma}) \circ
\hat{G} =  \hat{1},\quad \hat{G} \circ (\hat{G}^{-1}_0 - \hat{\Sigma}) =  \hat{1}$, we obtain  the evolution of the spectral sector

\begin{align}
    (i\partial_t - \xi_k) & G^R_{t t'}(k,k') + \frac{i\gamma}{2N}\theta(t-t_0) \\
   &\times \sum_q G^R_{t t'}(q,k') = \delta(t-t')\delta_{kk'} ,\notag{}\\
    (-i\partial_{t'} - \xi_{k'}) & G^R_{t t'}(k,k') + \frac{i\gamma}{2N}\theta(t'-t_0)\\
    & \times \sum_q G^R_{t t'}(k,q) = \delta(t-t')\delta_{kk'}.\notag{}
\end{align}
Interestingly, these equations have the same form as the ones for the single-impurity scattering problem, provided the strength of the potential is imaginary.

Similarly, we obtain the equations for the Keldysh sector (we now fix $t,t' \geq t_0$):
\begin{align}
    (i\partial_t - \xi_k) & G^K_{t t'}(k,k') + \frac{i\gamma}{2N} \sum_q G^K_{t t'}(q,k')  \\
    & =\frac{\gamma}{N^2} \sum_{p,q} G^K_{t t}(p,q) \times \sum_{q'}  G^A_{t t'}(q',k'),\notag{}\\
    (-i\partial_{t'} - \xi_{k'}) & G^K_{t t'}(k,k') - \frac{i\gamma}{2N}\sum_q G^K_{t t'}(k,q) \\
    & =\frac{\gamma}{N^2} \sum_{p,q} G^K_{t' t'}(p,q) \times \sum_{q'}  G^R_{t t'}(k,q'). \notag{}
\end{align}
These two equations can be combined into a single one on the equal-time Keldysh Green's function. Using the regularisation discussed above when $t\to t'$, we find $G^{R,A}_{tt}(k,k') = \mp\frac{i}{2}\delta_{kk'}$, and we arrive at
\begin{align}
    \left(i\frac{d}{dt} -(\xi_k-\xi_{k'})\right) & G^K_{t t}(k,k') = \frac{\gamma i}{N^2} \sum_{p,q} G^K_{t t}(p,q)   \\
    & -\frac{\gamma i}{2N}\sum_q \left(G^K_{t t}(q,k')+G^K_{t t}(k,q)  \right),\notag{}
\end{align}
which is Eq.~(3) in the main text.\\

\section{Non-equilibrium diagrammatic approach: NESS}
\subsection{Retarded sector}

Using our result for the retarded self-energy~\eqref{eqn::Sigma_R_SCBA}, we find that the steady-state retarded Green's function satisfies the following equation:
\begin{align}
G^R_\omega(x,x') = G^{R}_{0,\omega}(x,x') - \frac{i\gamma}{2} G^{R}_{0,\omega}(x,0)G^R_\omega(0,x'), \label{eqn::stationary_R}
\end{align}
which, in turn, is easy to solve:
\begin{align}
    G^R_\omega(x,x') = G^{R}_{0,\omega}(x,x') -\frac{i\gamma}{2} \frac{G^{R}_{0,\omega}(x,0)G^{R}_{0,\omega}(0,x')}{1 +\frac{i\gamma}{2} G^{R}_{0,\omega}(0,0)}.
\end{align}
An analogous result holds for the loss channel~[\onlinecite{froml2019fluctuation}].

\subsection{Keldysh sector}

Likewise, using the result for the Keldysh self-energy~\eqref{eqn::Sigma_K_SCBA}, we rewrite the Dyson equation as:
\begin{align}
    & G^K_{\omega}(x,x')  =   G^K_{0,\omega}(x,x') - \frac{i\gamma}{2}  G^R_{0,\omega}(x,0) G^K_{\omega}(0,x') \label{eqn::stationary_K}  \\
    &\quad + \gamma {\cal K} G^R_{0,\omega}(x,0) G^A_{\omega}(0,x') + \frac{i\gamma}{2}  G^K_{0,\omega}(x,0) G^A_{\omega}(0,x'),\notag{}
\end{align}
where ${\cal K} \equiv \int \frac{d\omega}{2\pi} G^K_{\omega}(0,0)$ is the equal-time correlation function at the origin. Assuming that ${\cal K}$ is a known constant, Eq.~\eqref{eqn::stationary_K} can be formally solved:
\begin{align}
    G^K_\omega(x,x') = F_{\omega}(x,x') -\frac{i\gamma}{2} \frac{G^{R}_{0,\omega}(x,0)F_{\omega}(0,x')}{1 +\frac{i\gamma}{2} G^{R}_{0,\omega}(0,0)},
\end{align}
where 
\begin{align}
    F_{\omega}(x,x')  \equiv & G^K_{0,\omega}(x,x') + \gamma {\cal K} G^R_{0,\omega}(x,0) G^A_{\omega}(0,x') \notag{} \\
    &\qquad \qquad + \frac{i\gamma}{2}  G^K_{0,\omega}(x,0) G^A_{\omega}(0,x').\notag{}
\end{align}
Substituting the results of the previous subsection, we arrive at:
\begin{widetext}
\begin{align}
    G^K_\omega(x,x') & = G^K_{0,\omega}(x,x') + \frac{i\gamma}{2}\frac{G^K_{0,\omega}(x,0)G^A_{0,\omega}(0,x')}{1 -\frac{i\gamma}{2} G^{A}_{0,\omega}(0,0)}- \frac{i\gamma}{2}\frac{G^R_{0,\omega}(x,0)G^K_{0,\omega}(0,x')}{1 +\frac{i\gamma}{2} G^{R}_{0,\omega}(0,0)} \notag{}\\
    &+ \frac{\gamma^2 G^K_{0,\omega}(0,0)}{4} \frac{G^R_{0,\omega}(x,0)G^A_{0,\omega}(0,x')}{(1 +\frac{i\gamma}{2} G^{R}_{0,\omega}(0,0))(1 -\frac{i\gamma}{2} G^{A}_{0,\omega}(0,0))} + \gamma {\cal K}\frac{G^R_{0,\omega}(x,0)G^A_{0,\omega}(0,x')}{(1 +\frac{i\gamma}{2} G^{R}_{0,\omega}(0,0))(1 -\frac{i\gamma}{2} G^{A}_{0,\omega}(0,0))}.\label{eqn::G_K_NESS}
\end{align}
From this, we obtain for $x = x' = 0$:
\begin{align}
    G^K_\omega(0,0) = G^K_{0,\omega}(0,0) \left| \frac{1}{1 +\frac{i\gamma}{2} G^{R}_{0,\omega}(0,0)} \right|^2 + \gamma {\cal K} \left| \frac{G^{R}_{0,\omega}(0,0)}{1 +\frac{i\gamma}{2} G^{R}_{0,\omega}(0,0)} \right|^2.
\end{align}
By integrating this equation over frequencies, we calculate the constant ${\cal K}$:
\begin{align}
    {\cal K} = \int \frac{d\omega}{2\pi} G^K_{0,\omega}(0,0) \left| \frac{1}{1 +\frac{i\gamma}{2} G^{R}_{0,\omega}(0,0)} \right|^2 \times \left[ 1 - \gamma \int \frac{d\omega}{2\pi} \left| \frac{G^{R}_{0,\omega}(0,0)}{1 +\frac{i\gamma}{2} G^{R}_{0,\omega}(0,0)} \right|^2 \right]^{-1}. \label{eqn::K}
\end{align}
Therefore, equations~\eqref{eqn::G_K_NESS} and~\eqref{eqn::K} give the full solution of the frequency-resolved Keldysh Green's function in the NESS.

\section{One-dimensional tight-binding model}

For the tight-binding model in the thermodynamic limit ($N\to\infty$), the free retarted Green's function is given by:
\begin{eqnarray}
G^R_{0,\omega} (x,x') = \frac{-i}{\sqrt{4t^2 - (\omega+\mu)^2}} \left[-\frac{\omega +\mu}{2J} + i\sqrt{ 1 - \left(\frac{\omega +\mu}{2J}\right)^2} \right]^{|x-x'|}.
\end{eqnarray}
 From the latter, we find
\begin{align}
\gamma \int \frac{d\omega}{2\pi} \left| \frac{G^{R}_{0,\omega}(0,0)}{1 +\frac{i\gamma}{2} G^{R}_{0,\omega}(0,0)} \right|^2  = \frac{2\alpha}{\pi} \left[ \int\limits_0^1 \frac{dx}{(\sqrt{1-x^2} + \alpha )^2 } + \int\limits_1^\infty \frac{dx}{x^2 - 1 + \alpha^2} \right],
\end{align}
where $\alpha \equiv \gamma/4J$ is the dimensionless heating strength. We calculate 
  \begin{equation}
   \int\limits_1^\infty \frac{dx}{x^2 - 1 + \alpha^2} = 
    \begin{cases*}
      \frac{\tanh^{-1}\sqrt{1-\alpha^2} }{\sqrt{1-\alpha^2}} & if $\alpha < 1$ \\
      1 & $\alpha = 1$ \\
      \frac{\tan^{-1}\sqrt{\alpha^2-1} }{\sqrt{\alpha^2-1}} & if $\alpha > 1$
    \end{cases*},\quad
      \int\limits_0^1 \frac{dx}{(\sqrt{1-x^2} + \alpha )^2 } =
          \begin{cases*}
      \frac{2}{1-\alpha^2} \frac{\tanh^{-1}\sqrt{\frac{1-\alpha}{1+\alpha}} }{\sqrt{1-\alpha^2}} - \frac{1}{1-\alpha^2}& if $\alpha < 1$ \\
      \frac{1}{3} & $\alpha = 1$ \\
      -\frac{2}{\alpha^2-1} \frac{\tan^{-1}\sqrt{\frac{\alpha-1}{1+\alpha}} }{\sqrt{\alpha^2-1}} + \frac{1}{\alpha^2-1} & if $\alpha > 1$
    \end{cases*},
  \end{equation}
  and therefore we get 
  \begin{align}
      \int \frac{d\omega}{2\pi} G^K_{0,\omega}(0,0) \left| \frac{1}{1 +\frac{i\gamma}{2} G^{R}_{0,\omega}(0,0)} \right|^2 = \frac{2i}{\pi} \int\limits_0^{\tilde{\mu}} \frac{\sqrt{1-x^2}dx}{(\alpha+\sqrt{1-x^2})^2},
  \end{align}
  where $\tilde{\mu}\equiv \mu/2J$ is the rescaled chemical potential. The latter integral can be evaluated analytically
  \begin{align}
      \int\limits_0^{\tilde{\mu}} \frac{\sqrt{1-x^2}dx}{(\alpha+\sqrt{1-x^2})^2} = 
      \begin{cases*}
      \arcsin{\tilde{\mu}} + \frac{\alpha^2}{1-\alpha^2}\frac{\tilde{\mu}}{\alpha+\sqrt{1-\tilde{\mu}^2}} - \frac{2\alpha(2-\alpha^2)}{(1-\alpha^2 )^{3/2}} \tanh^{-1}\left(\sqrt{\frac{1-\alpha}{1+\alpha}} \frac{\tilde{\mu}}{1 + \sqrt{1-\tilde{\mu}^2}}\right) & if $\alpha < 1$ \\
      \frac{3\arcsin{\tilde{\mu}} (2-\tilde{\mu}^2 + 2\sqrt{1-\tilde{\mu}^2}) - \tilde{\mu}(5\sqrt{1-\tilde{\mu}^2} +4)}{3(1+\sqrt{1-\tilde{\mu}^2})^2} & $\alpha = 1$ \\
      \arcsin{\tilde{\mu}} + \frac{\alpha^2}{1-\alpha^2}\frac{\tilde{\mu}}{\alpha+\sqrt{1-\tilde{\mu}^2}} - \frac{2\alpha(\alpha^2-2)}{(\alpha^2 -1 )^{3/2}} \tan^{-1}\left(\sqrt{\frac{\alpha-1}{\alpha+1}} \frac{\tilde{\mu}}{1 + \sqrt{1-\tilde{\mu}^2}}\right) \ & if $\alpha > 1$
    \end{cases*}.
  \end{align}

  \subsection{Density profile in the NESS}

Using the previous results, we obtain the density profile
\begin{align}
n_x & = \frac{1}{2}\left( 1 + \frac{2}{\pi} \arcsin{\tilde{\mu}} \right) - \frac{2\alpha}{\pi} \int\limits_{\arccos{\tilde{\mu}}}^{\pi/2} \frac{\cos^2(tx) dt }{\sin{t} + \alpha} + \frac{\alpha^2}{\pi} \int\limits_0^{\arcsin{\tilde{\mu}}} \frac{dt }{(\cos t + \alpha)^2} \notag{}\\
&\qquad + (2n_0 -1)\alpha/\pi \left[ \int\limits_0^1 \frac{dt} {(\alpha + \sqrt{1-t^2})^2} + \int\limits_1^\infty \frac{dt (\sqrt{t^2-1} - t )^{2|x|} }{\alpha^2 + (t^2-1)} \right]. \label{eqn::NESS_dens}
\end{align}
 with the initial state  given by the filled Fermi sea.
The average density for the NESS is different from the initial average density, which might be surprising given that the equations of motion respect the conservation of the total number of particles (there is no contradiction since the order of limits is crucial: first, we take $N\to \infty$ and then $t\to \infty$). It illustrates that even though the impurity is local, it affects the whole system up to spatial infinity (likewise, the case of static impurity). From this solution, one finds that there are two long-range contributions to the density at large distances: one which scales as $1/|x|$ and the other one scaling as $1/x^2$.
\end{widetext}

\section{Dynamics towards the NESS}

\subsection{Short-time dynamics}

It is instructive to investigate the two-site limit of the main model, which captures the short-time dynamics for strong dissipation. The effective equation of motion reads
\begin{align}\label{eq:spin}
 \frac{d}{dt}{\bf S} = {\bf B}\times {\bf S} - \frac{\gamma}{2} (S_x, S_y, 0)^T
\end{align}
with ${\bf S} =\frac{1}{2}\langle\hat{c}_i^\dagger {\bf \sigma}_{ij} \hat{c}_j
\rangle, i,j = 1,2$ a two-level spin system, describing the two sites with one of them  subject to the noisy perturbation (second term in~\eqref{eq:spin}). The `magnetic field' ${\bf B} = (-2J,~0,~0)^T$ encodes the hopping between the sites. When $\gamma \gg 4J$, the dynamics is governed by two eigenvalues: $\lambda_1 \approx - \gamma/2$ and $\lambda_2 \approx - 8 J^2/\gamma$. The second mode $\lambda_2$ is parametrically small, implying a `trapping' of particles jumping on the impurity site, since  strong noise  drives out of resonance hopping processes involving the dissipative site. This simple argument captures the short-time dynamics for strong dissipation, as shown in the inset of Fig. 2c. In addition, the initial decay rate extracted from this inset is fitted by a $1/\gamma$ dependence confirming the validity of this heuristic picture and agreeing with the  intuition about the onset of a Zeno effect.

\subsection{Approach to the NESS}
One can derive a self-consistent integral equation solely on the dynamics of $n_{0}(t)$:
\begin{align}
    &n_{0}(t)  = n_{\rm loss}(t) + \gamma \int_0^t dt' \, \Big| S_0(t-t') \Big|^2 n_{0}(t'),\label{eqn::n_0_integral}
\end{align}
where
\begin{align}
    &S_{n}(t)  \equiv i^{|n|}J_{|n|}(2t)  \\
    & \quad -\frac{\gamma i^{|n|}}{2} \int\limits^t_0 du \,{\rm e}^{-\gamma u/2}\, \left( \frac{t-u}{t+u} \right)^{|n|/2} J_{|n|}(2\sqrt{t^2-u^2})\notag{}
\end{align}
is obtained from the retarded Green's function~\eqref{eqn::stationary_R} by performing a Fourier (Laplace) transform~[\onlinecite{krapivsky2019free}]. $n_{\rm loss}$ represents the evolution of the density at origin for the loss channel, cf. Eq.~\eqref{eqn::losses}. We find that
\begin{align}
n_{\rm loss}(t) \equiv \sum_{m,n} S_m(t)S^*_n(t) n_{n,m}^{(0)}, \label{eqn::n_loss}
\end{align}
where $n_{n,m}^{(0)} = \frac{1}{\pi} \frac{\sin(k_F(m-n))}{m-n}$ represent the Fermi-sea correlations. 

Importantly, for the loss channel, the explicit sum~\eqref{eqn::n_loss} can be reliably computed, and we numerically find that $n_{\rm loss}(t)\simeq {\rm const} + At^{-2}$ (up to oscillatory behavior) at long times. This, together with the fact that the second term in Eq.~\eqref{eqn::n_0_integral} scales as $|S_{0}(t)|^2\sim t^{-3}$ at long times, allows to deduce that $(n_{0}(t)-n^{\infty}_{0})$ exhibits the  $t^{-2}$ scaling behavior discussed in the main text. 

\clearpage
\newpage

\section{Equation of motion for the 4-point correlation function}

\subsection{Local dephasing}

Using the formalism of the main text, we derive that $\Gamma^{k_1k_2}_{k_1'k_2'} \equiv \langle \hat{c}^\dagger_{k_1}\hat{c}^\dagger_{k_2}\hat{c}_{k_1'}\hat{c}_{k_2'}\rangle$ evolves according to
\begin{align}
    \frac{d\Gamma^{k_1k_2}_{k_1'k_2'} }{dt} & =  i(\xi_{k_1}+\xi_{k_2}-\xi_{k_1'}-\xi_{k_2'})\Gamma^{k_1k_2}_{k_1'k_2'}  \notag\\
    &- \frac{\gamma}{2N} \sum_q \left(\Gamma^{qk_2}_{k_1'k_2'} + \Gamma^{k_1q}_{k_1'k_2'} + \Gamma^{k_1k_2}_{qk_2'} + \Gamma^{k_1k_2}_{k_1'q} \right) \notag \\
    & + \frac{\gamma}{N^2} \sum_{pq}\left(\Gamma^{pk_2}_{qk_2'} + \Gamma^{pk_2}_{k_1'q} + \Gamma^{k_1p}_{qk_2'}+ \Gamma^{k_1p}_{k_1'q} \right).\label{eqn::dt_4point}
\end{align}
By writing the equation of motion for $$\tilde{\Gamma}^{k_1k_2}_{k_1'k_2'} \equiv \Gamma_{k_1k_2'}\Gamma_{k_2k_1'}-\Gamma_{k_1k_1'}\Gamma_{k_2k_2'}$$ one can verify that $\tilde{\Gamma}^{k_1k_2}_{k_1'k_2'}$  {\it will not} be a solution of Eq.~\eqref{eqn::dt_4point}, and, thus, the state is, in general, non-Gaussian.
In particular, the third term in Eq.~\eqref{eqn::dt_4point} is recognized to be responsible for this non-Gaussianity. 

\subsection{Local losses}

It is instructive to check that the above logic applied for the loss channel will show that dynamics are Gaussian. Following the same steps as for the dephasing, we derive the following equations:
\begin{align}
\frac{d}{dt}\Gamma_{k k'}(t) &= i(\xi_k - \xi_{k'})\Gamma_{k k'}
-\frac{\gamma}{2N} \sum_{q}(\Gamma_{k,q}+\Gamma_{q,k'}), \label{eqn::losses}\\
    \frac{d\Gamma^{k_1k_2}_{k_1'k_2'} }{dt} & =  i(\xi_{k_1}+\xi_{k_2}-\xi_{k_1'}-\xi_{k_2'})\Gamma^{k_1k_2}_{k_1'k_2'}  \notag\\
    &- \frac{\gamma}{2N} \sum_q \left(\Gamma^{qk_2}_{k_1'k_2'} + \Gamma^{k_1q}_{k_1'k_2'} + \Gamma^{k_1k_2}_{qk_2'} + \Gamma^{k_1k_2}_{k_1'q} \right)
    \label{eqn::dt_4point_losses}
\end{align}
From Eq.~\eqref{eqn::losses}, one can easily show that $\tilde{\Gamma}^{k_1k_2}_{k_1'k_2'}$ will indeed satisfy Eq.~\eqref{eqn::dt_4point_losses}.

\begin{figure}[t!]
\centering
\includegraphics[width=1\linewidth]{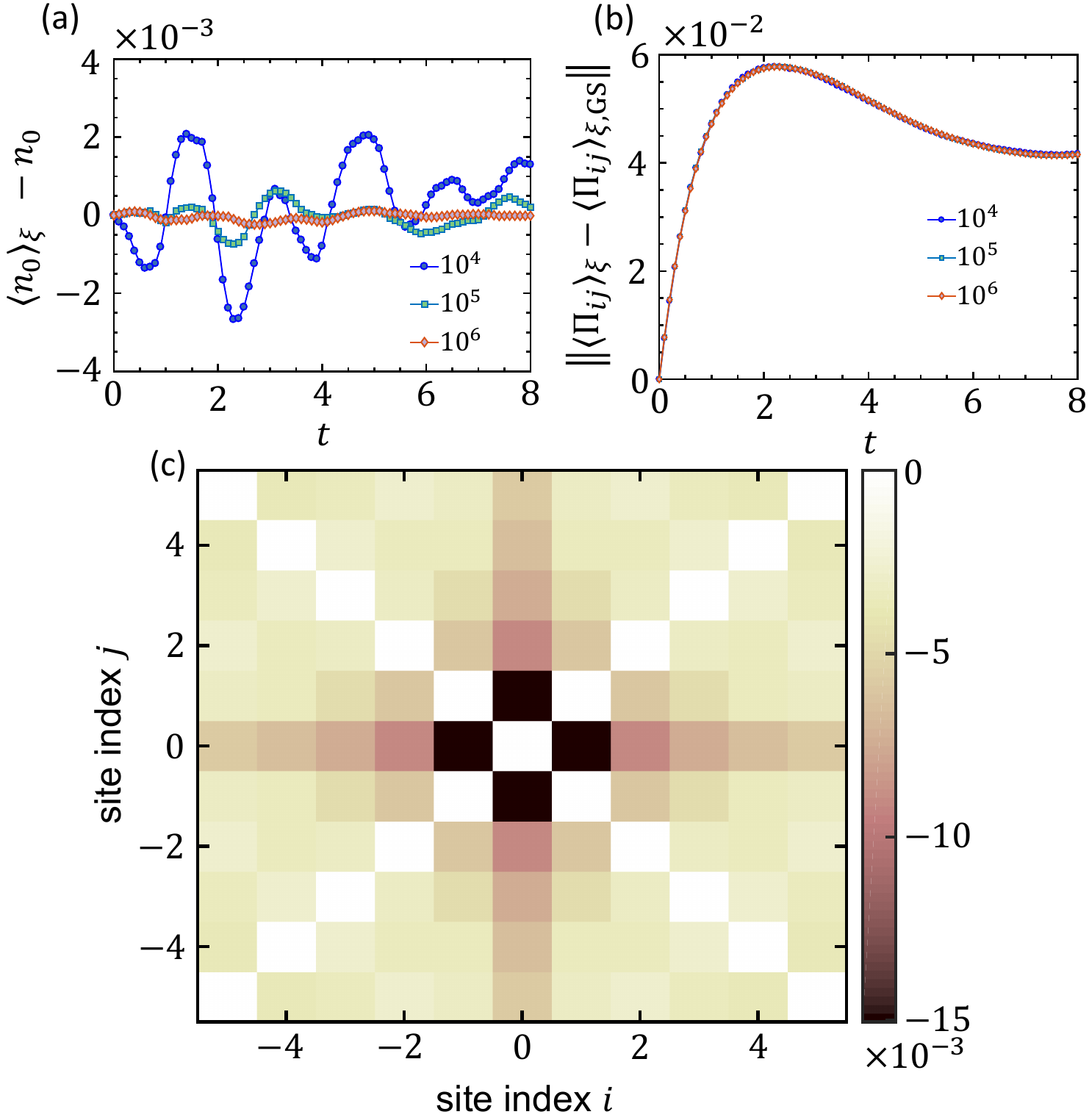} 
\caption{(a) Evolution of the noise-averaged   density at the origin, $\langle n_{0}(t) \rangle_\xi$, relative the one obtained from the QME, cf. Eq. (3). This plot shows that the two approaches, the SSE simulation and the Lindblad QME, start to match for sufficiently large noise statististics: for $N_\xi = 10^6$, the maximum error is about $2\times 10^{-4}$. (b) Evolution of $\norm{\langle\Pi_{ij}\rangle_\xi - \langle\Pi^{\rm GS}_{ij}\rangle_\xi}$, where $\Pi_{ij}$ is the polarization operator (see the text) in the vicinity of the dissipative impurity, and $\Pi^{\rm GS}_{ij}$ represents the same observable  obtained assuming the density matrix is Gaussian. The plot illustrates dynamical development of non-Gaussian correlations. (c) Spatial distribution of the non-Gaussian correlations, $\langle\Pi_{ij}\rangle_\xi - \langle\Pi^{\rm GS}_{ij}\rangle_\xi$, in the nearby of the origin at $t = 10 J^{-1}$ (we fix $N_\xi = 10^6$). Parameters used: $N = 20$, $k_F = \pi/4$, $\gamma = 1$. }
\label{fig::NGmap}
\end{figure}

\section{Non-gaussianity from the SSE}

Here we simulate the SSE for a small system size to further address the question of reliable noise statistics and non-Gaussianity of the many-body density matrix. 

We first benchmark our SSE simulations with the Lindbladian dynamics, cf. Eq.~(3) of the main text. Figure~\ref{fig::NGmap}a shows the evolution of the averaged over noise density at the origin relative to the same quantity obtained from simulating the Lindblad QME. We observe that the discrepancy between the two approaches is already small for $N_\xi = 10^4$, and it further decreases with increasing the total number of noise samples. We also note that this discrepancy does not increase with time, as it should be for bounded operators (such as fermionic density).

We now turn to discuss the four-point correlation function. For concreteness, we consider here the density-density correlation function (the polarization operator): $\Pi_{ij} \equiv \langle\hat{c}^\dagger_i\hat{c}^\dagger_j\hat{c}_j\hat{c}_i\rangle,$ $i,j = -l,\dots, l$ (we fix $l=5$). Figure~\ref{fig::NGmap}b demonstrates the evolution of the norm $\norm{\langle\Pi_{ij}\rangle_\xi - \langle\Pi^{\rm GS}_{ij}\rangle_\xi}$, where $\Pi^{\rm GS}_{ij}$ is the polarization matrix obtained assuming the state is Gaussian, i.e. from the two-body correlation function. This plot clearly demonstrates the devolopment of the non-Gaussian correlations. Notably, the three curves obtained for $N_\xi = 10^4, 10^5$, and $N_\xi = 10^6$, seem almost to coincide, suggesting that $N_\xi = 10^4$ already provides sufficient noise statistics to study the four-point correlation function. In Fig.~\ref{fig::NGmap}c, we plot the spatial spread of the non-Gaussian correlations, $\langle\Pi_{ij}\rangle_\xi - \langle\Pi^{\rm GS}_{ij}\rangle_\xi$, indicating that they are most pronounced near the site with decoherence. This feature persists for larger systems and for longer times.

\end{document}